\documentstyle[12pt]{article}

\newcommand{\eq}{\begin{equation}}
\newcommand{\en}{\end{equation}}
\newcommand{\eqn}{\begin{eqnarray}}
\newcommand{\enn}{\end{eqnarray}}
\newcommand{\nn}{\nonumber \\}
\newcommand{\barr}{\begin{array}}
\newcommand{\earr}{\end{array}}

\newcommand{\ft}[2]{{\textstyle {{#1}\over {#2}} }}

\newcommand{\NP}{{\it Nucl. Phys.}}
\newcommand{\PL}{{\it Phys.Lett.}}

\begin{document}
\begin{titlepage}
\begin{flushright}
PSU-TH-157 \\
January 1995
\end{flushright}
\vspace{1cm}
\begin{center}
{\bf Seven Dimensional Octonionic Yang-Mills Instanton and its Extension to an
Heterotic  String Soliton}\\

\vspace{1cm}
{\large Murat G\"{u}naydin} \\
Physics Department, 104 Davey Lab. \\
Penn State University, University Park PA16802, USA \\
\vspace{1cm}

{\large Hermann Nicolai} \\
II. Institut f\"{u}r Theoretische Physik \\
Universit\"{a}t Hamburg, Luruper Chaussee 149 \\
D-22761 Hamburg 50, Germany
\vspace{1cm}

{\bf Abstract}
\end{center}
We construct an octonionic instanton solution to the seven dimensional
 Yang-Mills
theory based on the exceptional gauge group $G_2$ which is the
 automorphism group
of the division algebra of octonions. This octonionic instanton
 has an extension to a solitonic solution of the low energy
 effective theory of the heterotic string that preserves
two of the sixteen supersymmetries and hence corresponds to $N=1$ space-time
supersymmetry in  $(2+1)$ dimensions transverse to
the  seven dimensions where the Yang-Mills instanton is defined.

\end{titlepage}

Recently there has been a resurgence of interest in solitonic solutions
 of various string theories, in particular,
 of the heterotic string \cite{CHS,DKL}.
 It is hoped that solitonic solutions will shed some
 light on the non-perturbative dynamics of string theories.
 Furthermore,
they are essential in verifying the validity of
 generalized duality conjectures in string theories \cite{duality}. 
In this letter our aim is to show that that there exists a solution
 to the self-duality equations in seven dimensional Yang-Mills theory
 with the gauge group $G_2$.
 The exceptional group $G_2$ is the automorphism group of the division algebra of octonions 
${\bf O}$ \cite{GG} and our solution depends in an intricate
 manner on the existence of this algebra. Hence the term ``octonionic
 instanton". Eight dimensional octonionic instantons based on the gauge group
 $Spin(7)$ were constructed in 
\cite{FaNu,FuNi}. To our knowledge the solution we give below is the only known
Yang-Mills instanton solution in odd dimensions.
 It is not difficult to convince oneself that
seven  is the unique odd dimension where such a solution can exist.
 In the second part of
this paper we extend the seven dimensional Yang-Mills instanton to a solitonic solution
of the low energy effective theory of the heterotic string ( to order
 $ \alpha' $) in parallel with the work of
of \cite{HaSt} where the eight-dimensional octonionic instanton was
extended to a soliton solution of the the heterotic string.

The exceptional group $G_2$  can be characterized as the maximal
common subgroup
of the seven dimensional rotation group $SO(7)$ and
its covering group $Spin(7)$ \cite{GG}. 
\eq
G_2 = SO(7) \bigcap Spin(7)
\en
For the purposes of this paper we shall
regard $G_2$ as  a subgroup of $Spin(7)$ taken in
its eight-dimensional spinor representation
 which decomposes as
\eq
{\bf 8} = {\bf 7} + {\bf 1}
\en
under its $G_2$ subgroup.
The $Spin(7)$ generators are given by
\eq
\Gamma^{mn} = \Gamma^{{[}m} \Gamma^{n{]}}
\en
where the $\gamma$ matrices in seven dimensions are defined as usual by 
\eq
\{ \Gamma^m , \Gamma^n \} = 2 \delta^{mn} 
\en
with $m,n,...= 1,2,...,7$.\footnote{ Throughout this paper the
 antisymmetrization ${[}m,n,...{]}$ of indices will be of weight one.}
The generators $G^{mn}= - G^{nm}$ of $G_2$ (considered as a subgroup of
$Spin(7)$) are defined through the constraints
\eq
C_{mnp} G^{np} = 0
\en
where $C_{mnp}$ are the structure constants of the octonion algebra
\cite{GG,DN}. Since the number of independent
 constraints is seven, there are altogether
14 generators of $G_2$.
In terms of the matrices $\Gamma^{mn}$ the $G_2$ generators
can be written as \cite{DN}
\eq
G^{mn} = \ft12 \Gamma^{mn} + \ft18 C_{mnpq} \Gamma^{pq}
\en
where the completely antisymmetric tensor $C^{mnpq}$ is defined as
\eq
C^{mnpq}:= \ft16 \epsilon^{mnpqrst} C_{rst}
\en
These tensors are subject to the identities
\eqn
C^{mnp} C_{np}^{~~qr} &=& -4 C^{mqr} \nn
C_{mnp} C^{qrsp} &=& 6 \, \delta^{{[}q}_{{[}m} C_{n{]}}^{~rs{]}} \\
C^{mnrs} C_{pqrs} &=& 8 \, \delta^{mn}_{pq} - 2 C^{mn}_{~~~pq} \nonumber
\enn
which can be derived by use of the properties of the structure
constants $C_{mnp}$ and the corresponding identities
in eight dimensions \cite{DN,DGT}.
The $G_2$ commutation relations
can be determined from the corresponding commutation relations of
$Spin(7)$ (cf. eq.(11) of \cite{FuNi}) 
by specializing to its $G_2$ subgroup. Using the above identities
we find after some calculation
\eq
{[}G^{mn} , G^{pq} {]} = 2 \delta^{p{[}n} G^{m{]}q} - 2 \delta^{q{[}n}
 G^{m{]}p} 
  + \ft12 \Big( C^{pqr{[}m} G^{n{]}r}
 - C^{mnr{[}p} G^{q{]}r}\Big)
\en
As a check on these relations, readers may verify that the right hand side
vanishes upon contraction with either $C_{smn}$ or $C_{spq}$.
The constraint (5) implies the identity 
\eq
G_{mn} = \ft12 C_{mnpq} G^{pq}
\en
For later use let us also record the normalization of these generators
\eq
{\rm Tr} \, G_{mn} G^{pq} = -9 {P_{mn}}^{pq}
\en
where ${P_{mn}}^{pq}:= \ft23 (\delta_{mn}^{pq} + \ft14 {C_{mn}}^{pq} )$
is the projector on the $G_2$ subalgebra of $Spin(7)$
(with $ \delta_{mn}^{pq} = \delta^{{[}p}_{{[}m} \delta^{q{]}}_{n{]}}$).

Consider now the Yang-Mills gauge
theory in seven dimensions with the gauge group 
$G_2$. In analogy with \cite{FuNi} we proceed from the following ansatz for the Yang-Mills gauge field $A_m (y)$:
\eq
A_m (y) = G_{mn} f_n (y)
\en
where $f_n (y) \equiv \partial_n f(y)$ and $f(y)$
 is a scalar function of the
coordinates $y^n$ to be determined by the self-duality condition.
 The field strength
\eq
F_{mn} = \partial_m A_n - \partial_n A_m + {[} A_m , A_n {]}
\en
corresponding to this ansatz  takes the form
\eqn
F_{mn} (y) &=& 2 f_{p{[}m} G_{n{]}p} - 2 f_q G_{q{[}m} f_{n{]}} \nn
& & - G_{mn} (f_q f_q ) -\ft12 C_{mnpq} f_r G_{rq} f_p 
\enn
where $f_{pq} \equiv \partial_p \partial_q f(y) $.
We define the dual field strength $\tilde{F}_{mn}
 $ using the $G_2 $ invariant tensor
$C_{mnpq} $:
\eq
\tilde{F}_{mn} (y) = \lambda C_{mnpq} F_{pq} (y)
\en
where $\lambda$ is a constant to be determined. We get
\eqn
\tilde{F}_{mn} =& \lambda \Big( 2 C_{mnst} f_{ps} G_{tp}
 - 3 C_{mnst} f_q G_{qs} f_t \nn
  & -2 (f_q f_q ) G_{mn} + 4 f_q G_{q{[}m} f_{n{]}} \Big)
\enn
To solve the self-duality condition
\eq
F_{mn} = \tilde{F}_{mn}
\en
we put $f(y) = -\frac{1}{2} {\rm log} \phi (y)$. 
 One finds that $\lambda = \frac{1}{2}$ and
\eq
\phi (y) = \rho^2 + r^2
\en
with $r^2 \equiv y^m y^m$.
Hence the gauge field for this instanton is simply
\eq
A_m = - \frac{G_{mn} y_n }{ \rho^2 + r^2 }
\en
where $\rho$ is an arbitrary scale parameter.

Let us now show that the above instanton solution can be extended to
 a solitonic solution
of the heterotic string.
 Consider the purely bosonic sector of the ten dimensional 
low energy effective theory of the heterotic string
 ( to order $\alpha'$).
\eq
S = \frac{1}{2 \kappa^2} \int d^{10}X \sqrt{-g}  e^{-2\phi(X)}
 \bigg( R +4 (\vec{\nabla}\phi)^2 -\ft13 H^2 -
 \ft1{30} \alpha' {\rm Tr} \, (F^2 ) \bigg)
 \en
where $X^M \;  (M=0,1,...,9)$ 
denote the coordinates of ten dimensional spacetime.
We are interested in solutions that preserve at
 least one supersymmetry. This requires that
in ten dimensions there exist at least one Majorana-Weyl
 spinor $\epsilon$ such that
the supersymmetry variations of the fermionic fields vanish for such
 solutions \cite{CHS}
\eqn
\delta \chi(X) &=& F_{MN} \Gamma^{MN} \epsilon = 0 \nn
\delta \lambda(X) & = & \big( \Gamma^M \partial_M \phi -
 \ft16  H_{MNP} \Gamma^{MNP} \big) \epsilon = 0  \\
\delta \psi_M (X) &=& \big( \partial_M + 
\ft14    \Omega_{-M}^{~~~AB} \Gamma_{AB} \big) \epsilon = 0  \nonumber
\enn
where $\chi$, $\lambda$ and $\psi_M $ are the gaugino, dilatino and the gravitino fields, respectively.  The generalized connection $\Omega_{-}$ is defined as
\eq
\Omega_{-M}^{~~AB} = \omega_{M}^{~AB} - H_{M}^{~AB} 
\en
where $\omega$ is the spin connection and
 $H$ is the anti-symmetric tensor field strength. We 
denote the ten dimensional world indices as $M,N,.. = 0,1,...,9$
 and the corresponding tangent space indices as $A,B,...= 0,1,...,9$.
 Since we are interested in solutions that extend the octonionic instanton
 we decompose the indices as
\eqn
M=(\alpha, \mu) &;& N=(\beta, \nu) ,...\nn
A=(a, m) &;& B=(b, n) , ...\\
\alpha , \beta, .. =0,1,2 &;& \mu, \nu, ... = 3,4,..,9 \nn
a,b,..=0,1,2  &;& m,n,...= 3,4,..,9 \nonumber
\enn
Note that the indices $m,n,...$ that were running from $1$ to $7$ now run
 from $3$ to $9$ so as to agree with the standard convention of denoting
 the timelike coordinate as $X^0$.
The coordinates $y^m$ of the instanton solution will  be 
identified with $X^m$ .
For the purposes of this paper we shall restrict ourselves to solutions
 that are Poincare
invariant in  $(2+1)$ dimensions. First we choose $\epsilon$ to be 
a $G_2$ singlet of
the Majorana-Weyl spinor of $SO(9,1)$.
 There are two such singlets since under
the $G_2 \times SO(2,1)$ subgroup of $SO(9,1)$ the Majorana-Weyl spinor 
decomposes as ${\bf 16} = ({\bf 1},{\bf 2}) + ({\bf 7},{\bf 2})$.
Let us denote these singlets as $\eta^i $
 ($i=1,2$).
Thus taking $\epsilon$ to be a $G_2$ singlet $\eta$ and
 the non-vanishing components of $F_{MN}$ to be those given by
 the seven dimensional octonionic instanton the supersymmetry variation
 $\delta \chi$ vanishes. This follows from the fact that
\eq
G^{mn} \eta =0
\en
and the self-duality of $F_{mn}$.
 The vanishing of the supersymmetry 
variation $\delta \lambda$ of the dilatino
 requires that  the non-vanishing components $H_{mnp} (y)$ of the antisymmetric
 tensor field strength be related to the dilaton $\phi$ as follows:
\eq
H_{mnp} =- \ft14 C_{mnpq} \partial^q \phi (y)
\en
With this choice of $H_{mnp}$ the gravitino variation $\delta \psi_M$
 also vanishes if we take
the metric  $g_{\mu\nu}$ 
in the seven dimensional subspace to be of the form
\eq
g_{\mu\nu} = e^{\phi(y)} \delta_{\mu\nu}  \hspace{1cm} (\mu, \nu =3,...,9)
\en
with a dilaton field $\phi(y)$ that is a function of $y^{\mu}$ .
 To solve for this function and thus the
dilaton field we need to further impose the Bianchi identity:
\eq
dH = \alpha' \Big({\rm tr} R \wedge R - \ft1{30}{\rm Tr} F \wedge F \Big)
\en
where Tr refers to the trace in the fundamental representation
 of $E_8$ or $SO(32)$ in the corresponding heterotic string theory.
 (For $E_8$ the fundamental representation 
coincides with the adjoint representation).
To order $\alpha'$ we can neglect the first term \cite{CHS} and we have
\eq
dH = - \ft1{30} \alpha' \, {\rm Tr} F \wedge F = - \alpha' \, {\rm Tr}_8 F \wedge F
\en
where ${\rm Tr}_8$ refers to the trace in the spinor
 representation of $Spin(7)$ that contains $G_2$. 
Using
\eq
{\rm Tr} \, F_{[mn} F_{pq]} = \partial_{[m} \Phi_{npq]}
\en
with
\eq
\Phi_{mnp} (y) := 2 \frac{(3 \rho^2 + r^2 )}{(\rho^2 + r^2)^3} C_{mnpq} y^q
\en
we find
\eq
e^{-\phi (y)} = e^{-\phi_0} +4 \alpha'
 \frac{(2 \rho^2 + r^2)}{(\rho^2 + r^2)^2 }
\en
where $\phi_0$ is the value of the dilaton in the limit
$r \rightarrow \infty$.

Total ADM mass per unit (d-1)-volume of a (d-1)-brane is given by 
\cite{DGHR,DKL}
\eq
M_{d}= \int d^{(10-d)}y \Theta_{00}
\en
where $\Theta_{MN}$ is the total energy-momentum pseudotensor of the 
combined gravity-matter system. The metric $g_{mn}(y)$ corresponding to
our solution is asymptotically flat with a $1/r^2$ falloff as 
$r \rightarrow \infty $. This leads to a divergent ADM mass per unit
two-brane volume just like the soliton of reference \cite{HaSt} corresponding
to the eight dimensional octonionic instanton. The authors of \cite{HaSt}
argue that this divergent energy is an infrared phenomenon and does not
preclude the existence of a well-behaved low-energy effective action
governing the string dynamics on scales large relative to its core size.
Their arguments are equally applicable to the soliton solution involving
the seven dimensional octonionic instanton we presented above. However,
the correct physical interpretation of  these solutions and its implications
for superstring theory remain to be understood fully.

{\it Acknowledgement: } One of us (M.G) would like to thank the II. Institut
f\"{u}r Theoretische Physik and DESY for their hospitality where part of this 
work was done.

\newpage

 \begin{center}
   {\large\bf Addendum}\\
   \vspace{1pc}
 
   {\Large\sf
   Seven-dimensional octonionic Yang-Mills instanton and \\
 its  extension to an heterotic string soliton }\\
 \vspace{1pc}
   {\large\bf Physics Letters B351 (1995) 169-172} \\
 \vspace{1pc}
   {\large Murat G\"unaydin, Hermann Nicolai} \\
February 1996\\
 \end{center}
 
Recently we have been informed that self-dual solutions of the
Yang-Mills equations for arbitrary gauge groups in dimensions
$d\geq 4$ were already described in
 \begin{enumerate}
 \item A.D.~Popov, Europhys. Lett. {\bf 17} (1992) 23-26
 \item T.A.~Ivanova and A.D.~Popov, Lett. Math. Phys. {\bf 24} (1992) 85-92
 \item T.A.~Ivanova and A.D.~Popov, Theor. Math. Phys. {\bf 94} (1993) 225-242
 \item T.A.~Ivanova, Phys. Lett. {\bf B315} (1993) 277-282
 \end{enumerate}
In the last paper the instanton solutions in seven and 
eight dimensions are extended to heterotic string solitons. 
While their construction is more general than ours, however,
the above authors do not explicitly study the instanton solution 
of Yang-Mills theory based on the exceptional gauge group $G_2$, 
which was given in our paper and extended to an heterotic
string 2-brane.
 \end{document}